\documentclass
[twocolumn,showpacs,preprintnumbers,nofootinbib,superscriptaddress]{revtex4}%
\usepackage{amsfonts}
\usepackage{amsmath}
\usepackage{amssymb}
\usepackage{graphicx}%
\begin{document}
\title{Remarks about the microcanonical description of astrophysical systems.}
\author{L. Velazquez}
\email{luisberis@geo.upr.edu.cu}
\affiliation{Departamento de F\'{\i}sica, Universidad de Pinar del R\'{\i}o, Mart\'{\i}
270, Esq. 27 de Noviembre, Pinar del R\'{\i}o, Cuba.}
\author{F. Guzm\'{a}n}
\email{guzman@info.isctn.edu.cu}
\affiliation{Departamento de F\'{\i}sica Nuclear, Instituto Superior de Ciencias y
Tecnolog\'{\i}a Nucleares, Carlos III y Luaces, Plaza, La Habana, Cuba.}
\date{\today}

\begin{abstract}
We reconsider some general aspects about the mean field thermodynamical
description of the astrophysical systems based on the microcanonical ensemble.
Starting from these basis, we devote a special attention to the analysis of
the scaling laws of the thermodynamical variables and potentials in the
thermodynamic limit. Geometrical considerations motivate a way by means of
which could be carried out a well-defined generalized canonical-like
description for this kind of systems, even being nonextensive. This
interesting possibility allows us to extend the applicability of the Standard
Thermodynamic methods, even in the cases in which the system exhibits a
negative specific heat. As example of application, we reconsider the classical
Antonov problem of the isothermal spheres.

\end{abstract}
\maketitle

\section{Introduction}

Traditional thermodynamics is not able to describe the astrophysical systems
because of their nonextensive nature: They are nonhomogeneous systems whose
total energy does not scale asymptotically with the particles number $N$
during the increasing of the system size, as well as they exhibit energetic
regions with a negative specific heat, which makes the canonical description
unsuitable for describing their thermodynamical properties
\cite{pad,kies,ruelle,lind,antonov,Lynden,Lynden1,thirring,gross,chava}.

An adequate thermodynamical description for astrophysical systems can be
carried out starting from microcanonical basis under certain conditions
\cite{gross,de vega}. This possibility is very attractive because the
microcanonical ensemble is the only well-defined statistical ensemble: since
dynamics is always well defined, and because microcanonical ensemble is a
\textquotedblleft dynamical\textquotedblright\ ensemble.

The aim of the present paper is to reconsider some general questions about
the\ thermodynamic formalism for astrophysical systems based on the
consideration of the microcanonical ensemble. The organization of the paper is
the following. Sec. II\ is devoted to review some aspects of the
microcanonical mean field description of the astrophysical systems in order to
provide a general background for the analysis of the scaling laws of the
thermodynamical variables and potentials. Some details about the Legendre
formalism are also presented. Implications of the geometrical aspects of the
microcanonical ensemble on the applicability of a \textit{generalized
canonical-like} description are discussed in Sec. III. Finally, in Sec. IV
some conclusions are drawn.

\section{Microcanonical description}

A direct microcanonical calculation in a selfgravitating system is an
formidable task which can be performed by using Montecarlo numerical
simulations (see for example in refs.\cite{de vega,gross1}). The statistical
description of \textit{N}-body selfgravitating system is usually carried out
by using a mean field approximation when the particles number $N$ is large
enough \cite{gross,de vega}. We will perform in this section a brief review
about some aspects of this microcanonical mean field formalism for the
astrophysical systems.

\subsection{The microcanonical mean field approximation}

Let us to consider an astrophysical Hamiltonian system composed by a huge
number $N$ of identical particles enclosed in a rigid container\ with a
characteristic linear dimension $L$. The consideration of this rigid container
is a regularization procedure for the thermodynamical description for this
kind of system which avoids the particles evaporation \cite{antonov}. Let us
also suppose that these particles interact among them by means of the gravity
and short-range interactions, as example, forces with an
electrostatic\ origin. The Hamiltonian of this system can be discomposed into
two terms as follows:%

\begin{equation}
H_{N}=h_{N}+V_{N}, \label{Ham}%
\end{equation}
where $V_{N}$ is the potential energy associated to the Newtonian
gravitational interaction, while $h_{N}$ contains the energy contribution of
the short-range interactions and the kinetic energy of the particles.

The entropy function for the selfgravitating system in the microcanonical
description is obtained from the Boltzmann principle:%

\begin{equation}
S_{B}=\ln W,
\end{equation}
where the microcanonical accessible volume $W\left(  E,N;L\right)  $ is given by:%

\begin{equation}
W\left(  E,N;L\right)  =Sp\left[  \delta\left(  E-H_{N}\right)  \right]
\delta\epsilon_{0,} \label{W}%
\end{equation}
being $\delta\epsilon_{0}$ a suitable energy constant which makes $W$
dimensionless. Quantum effects or/and the consideration of a natural size for
the particles will be taken into account in an implicit manner in the equation
(\ref{W}), which is a necessary consideration in order to avoid the
short-range singularity of the Newtonian interaction, and therefore, the
gravitational collapse \cite{antonov,Lynden}.

We are interested in describing the large $N$ limit. This aim can be carried
out by using the following procedure: we partition the physical space in cells
$\left\{  c_{\alpha}\right\}  $ whose characteristic linear dimensions are
much larger than the effective radio of the short-range interactions, but very
short in comparison with the characteristic linear dimension $L$ of the
container. This assumption supposes that this container is very large. We
denoted by $n_{\alpha}=n\left(  \mathbf{r}_{\alpha}\right)  $ the number of
particles inside the cell $c_{\alpha}$, being $\mathbf{r}_{\alpha}$ the
position of its center.

The short-range interactions among the particles belonging to different cells
can be neglected, and therefore, the term $h_{N}$ of the Hamiltonian
(\ref{Ham}) can be approximated by:%

\begin{equation}
h_{N}\simeq\underset{\alpha}{\sum}h_{n_{\alpha}},
\end{equation}
where $h_{n_{\alpha}}$ is the energetic contribution of\ the $n_{\alpha}$
particles inside the cell $c_{\alpha}$. Hereafter, we refer this energetic
contribution as the internal energy. The potential energy $V_{N}$ associated
to the gravitational interaction can be approximated as follows:%

\begin{equation}
V_{N}\longrightarrow V\left[  n\right]  =-\underset{\alpha>\beta}{\sum}%
\frac{Gm^{2}n_{a}n_{\beta}}{\left\vert \mathbf{r}_{\alpha}-\mathbf{r}_{\beta
}\right\vert }.
\end{equation}

Taking into consideration all approximations exposed above, the microcanonical
volume $W$ (\ref{W}) can be rewritten as follows:%

\begin{equation}
W\left(  E,N,L\right)  \simeq\underset{\left\{  n_{\alpha}\right\}  }{\sum
}\delta_{D}\left(  N-\underset{\alpha}{\sum}n_{a}\right)  W\left[
n;E,L\right]  ,
\end{equation}
being:%

\begin{equation}
\delta_{D}\left(  k\right)  =\left\{
\begin{tabular}
[c]{ll}%
$1,$ & if $k=0$\\
$0$ & otherwise
\end{tabular}
\ \ \ \right.  \text{ and }\underset{\left\{  n_{\alpha}\right\}  }{\sum
}\equiv\underset{n_{1}}{\sum}\underset{n_{2}}{\sum}\cdots.
\end{equation}
The functional $W\left[  n;E,L\right]  $ is given by:%

\begin{align}
W\left[  n;E,L\right]   &  =\int\left(  \underset{\alpha}{\prod}\frac{de_{a}%
}{\delta\epsilon_{\alpha}}\right)  \delta\epsilon_{0}\delta\left(  E-V\left[
n\right]  -\underset{\beta}{\sum}e_{\beta}\right)  \times\nonumber\\
&  \times\exp\left[  \underset{\alpha}{\sum}S_{\alpha}\left(  e_{a},n_{\alpha
},v_{\alpha}\right)  \right]  ,
\end{align}
where $S_{\alpha}\left(  e_{\alpha},n_{\alpha},v_{\alpha}\right)  =\ln\left[
\omega_{\alpha}\left(  e_{\alpha},n_{\alpha},v_{\alpha}\right)  \right]  $ is
the Boltzmann entropy associated to the subsystem of $n_{\alpha}$ particles
inside the cell $c_{\alpha}$ whose internal energy is $e_{\alpha}$, being
$v_{\alpha}$ the volume of this cell. The quantity $\omega_{\alpha}\left(
e_{\alpha},n_{\alpha},v_{\alpha}\right)  $ is the number of microstates of
this subsystem with the above macroscopic configuration:%

\begin{equation}
\omega_{\alpha}\left(  e_{\alpha},n_{\alpha},v_{\alpha}\right)  =Sp\left(
\delta\left[  e_{\alpha}-h_{n_{\alpha}}\right]  \right)  \delta\epsilon
_{\alpha},
\end{equation}
where\ $\delta\epsilon_{a}$ is a suitable energy constant. Since
$h_{n_{\alpha}}$ involves only short-range interactions, this subsystem will
look-like as an extensive system for $n_{\alpha}$ large, and therefore, the
entropy per volume $s_{\alpha}$ depends only on the internal energy density
$\epsilon_{\alpha}=e_{\alpha}/v_{\alpha}$ and the particles density
$\rho_{\alpha}=n_{\alpha}/v_{\alpha}$ as follows:%

\begin{equation}
s_{\alpha}=S_{\alpha}\left(  e_{\alpha},n_{\alpha},v_{\alpha}\right)
/v_{\alpha}=s\left(  \epsilon_{\alpha},\rho_{\alpha}\right)  .
\end{equation}

The microcanonical volume $W$ can be conveniently rewritten by using the
following \textit{mean field approximation}:%

\begin{align}
W  &  \rightarrow W_{MF}\left(  E,N,L\right)  =A\int\mathcal{D}\rho\left(
\mathbf{r}\right)  \mathcal{D}\epsilon\left(  \mathbf{r}\right)  \exp\left(
S\left[  \epsilon,\rho\right]  \right)  \times\nonumber\\
&  \times\delta\left(  N-N\left[  \rho\right]  \right)  \delta\left(
E-H\left[  \epsilon,\rho\right]  \right)  , \label{MF}%
\end{align}
where $A$ is an unimportant constant which involves the constants
$\delta\epsilon_{\alpha}$'s and $v_{\alpha}$'s. The functionals%

\begin{equation}
S\left[  \epsilon,\rho\right]  =\int_{\mathbb{R}^{3}}d^{3}\mathbf{r}~s\left\{
\epsilon\left(  \mathbf{r}\right)  ,\rho\left(  \mathbf{r}\right)  \right\}  ,
\end{equation}

\begin{equation}
H\left[  \epsilon,\rho,\phi\left[  \rho\right]  \right]  =\int_{\mathbb{R}%
^{3}}d^{3}\mathbf{r}~\epsilon\left(  \mathbf{r}\right)  +\frac{1}{2}%
m\rho\left(  \mathbf{r}\right)  \phi\left(  \mathbf{r}\right)  ,
\end{equation}

\begin{equation}
N\left[  \rho\right]  =\int_{\mathbb{R}^{3}}d^{3}\mathbf{r}~\rho\left(
\mathbf{r}\right)  ,
\end{equation}
represent the total entropy, energy and particles number for a given profile
with $\epsilon\left(  \mathbf{r}\right)  $ and $\rho\left(  \mathbf{r}\right)
$, being $\phi\left(  \mathbf{r}\right)  $ the Newtonian potential:%

\begin{equation}
\phi\left(  \mathbf{r}\right)  =\mathcal{G}\left[  \rho;\mathbf{r}\right]
=-\int_{\mathbb{R}^{3}}\frac{Gm\rho\left(  \mathbf{r}_{1}\right)
d^{3}\mathbf{r}_{1}}{\left\vert \mathbf{r}-\mathbf{r}_{1}\right\vert },
\label{green}%
\end{equation}
where $\mathcal{G}\left[  \rho;\mathbf{r}\right]  $ is the Green solution of
the Poisson problem:%

\begin{equation}
\Delta\phi=4\pi Gm\rho.
\end{equation}

The expression of the microcanonical volume in mean field approximation can be
rewritten in order to avoid the $\rho$ dependence of the Newtonian potential
as follows:%

\[
W_{MF}\left(  E,N,L\right)  =A\int\mathcal{D}\rho\left(  \mathbf{r}\right)
\mathcal{D}\epsilon\left(  \mathbf{r}\right)  \mathcal{D}\phi\left(
\mathbf{r}\right)  \exp\left(  S\left[  \epsilon,\rho\right]  \right)  \times
\]

\begin{equation}
\times\delta\left\{  \phi\left(  \mathbf{r}\right)  -\mathcal{G}\left[
\rho;\mathbf{r}\right]  \right\}  \delta\left(  N-N\left[  \rho\right]
\right)  \delta\left(  E-H\left[  \epsilon,\rho,\phi\right]  \right)  .
\end{equation}
$W_{MF}\left(  E,N,L\right)  $ can be rewritten again by using the Fourier
representation of the delta functions, yielding:%

\begin{align}
W_{MF}\left(  E,N\right)   &  \sim\int_{-\infty}^{+\infty}\int_{-\infty
}^{+\infty}\frac{dkd\eta}{\left(  2\pi\right)  ^{2}}\int\mathcal{D}%
\rho\mathcal{D}\epsilon\mathcal{D}\phi\mathcal{D}h\times\nonumber\\
&  \times\exp\left\{  \mathcal{L}\left[  \epsilon,\rho,\phi;z,z_{1},J\right]
\right\}  , \label{wm}%
\end{align}
where $z=\beta+ik$ and $z_{1}=\mu+i\eta$ with $\beta,\eta\in\mathbb{R}$, being
the functional $\mathcal{L}\left[  \epsilon,\rho,\phi;z,z_{1},J\right]  $
defined by:%

\begin{equation}%
\begin{array}
[c]{c}%
\mathcal{L}\left[  \epsilon,\rho,\phi;z,z_{1},J\right]  =S\left[
\epsilon,\rho\right]  +z\left(  E-H\left[  \epsilon,\rho,\phi\right]  \right)
+\\
+z_{1}\left(  N-N\left[  \rho\right]  \right)  +J\cdot\left(  \phi
-\mathcal{G}\left[  \rho\right]  \right)  .
\end{array}
\label{Leg}%
\end{equation}
The functional term $J\cdot\left(  \phi-\mathcal{G}\left[  \rho\right]
\right)  =\int_{\mathbb{R}^{3}}d^{3}\mathbf{r}~J\left(  \mathbf{r}\right)
\left\{  \phi\left(  \mathbf{r}\right)  -\mathcal{G}\left[  \rho
;\mathbf{r}\right]  \right\}  $ appears as consequence of the Fourier
representation of the delta functional $\delta\left\{  \phi\left(
\mathbf{r}\right)  -\mathcal{G}\left[  \rho;\mathbf{r}\right]  \right\}  $.
Here, $J\left(  \mathbf{r}\right)  $ is a complex function, $J\left(
\mathbf{r}\right)  =j\left(  \mathbf{r}\right)  +ih\left(  \mathbf{r}\right)
$, with $j\left(  \mathbf{r}\right)  \in\mathbb{R}$.

\subsection{Scaling properties of the thermodynamical variables and
potentials}

The integration of expression (\ref{wm}) is usually carried out by using the
\textit{steepest decent method}. The application of this method is based on
the asymptotic behavior of the thermodynamical variables and potentials in the
many particle limit $N\gg1$.

The presence of an additive kinetic part in the Hamiltonian of certain system
leads to an exponential growing of the microcanonical volume $W$ with the $N$
increasing, and therefore, the Boltzmann entropy will grow proportional to $N$
in the many particles limit, $S_{B}=\ln W\propto N$.

The usual thermodynamic limit for the extensive systems:%

\begin{equation}
N\rightarrow\infty,\text{ keeping constant }\frac{E}{N}\text{ and }\frac{N}%
{V},
\end{equation}
where $V$ is the volume of the system, is directly related with the
\textit{extensive properties} of these systems when the thermodynamical
variables of the system are scaled by some scaling parameter $\alpha$ as follows:%

\begin{equation}
\left.
\begin{array}
[c]{c}%
N\rightarrow N\left(  \alpha\right)  =\alpha N_{,}\\
E\rightarrow E\left(  \alpha\right)  =\alpha E,\\
V\rightarrow V\left(  \alpha\right)  =\alpha V,
\end{array}
\right\}  \Rightarrow W\rightarrow W\left(  \alpha\right)  =\exp\left(
\alpha\ln W\right)  . \label{extensive}%
\end{equation}
In analogy with the extensive properties of the traditional systems, we will
analyze the necessary conditions for the existence of the following
\textit{power law self-similarity }scaling behavior of the microcanonical
variables $E$, $N$ and $L$ for the astrophysical systems:%

\begin{equation}
\left.
\begin{array}
[c]{c}%
N\rightarrow N\left(  \alpha\right)  =\alpha N_{,}\\
E\rightarrow E\left(  \alpha\right)  =\alpha^{\Lambda_{E}}E,\\
L\rightarrow L\left(  \alpha\right)  =\alpha^{\Lambda_{L}}L,
\end{array}
\right\}  \Rightarrow W\rightarrow W\left(  \alpha\right)  =\exp\left(
\alpha\ln W\right)  , \label{self}%
\end{equation}
where $\Lambda_{E}$ and $\Lambda_{L}$ are certain constant scaling exponent
which lead to an \textit{extensive }character of the Boltzmann entropy. This
kind of self-similarity behavior is directly related with a thermodynamic
limit of the form:%

\begin{equation}
N\rightarrow\infty\text{, keeping constant }\frac{E}{N^{\Lambda_{E}}}\text{
and }\frac{L}{N^{\Lambda_{L}}}. \label{THL}%
\end{equation}

The existence of this kind of self-similarity condition allows a considerable
simplification of the thermodynamical description: the study can be performed
by setting $N=1$ and considering the \textit{N}-dependence in the scaling laws
by taking $\alpha=N$. This scaling behavior is very useful in numerical
experiments, since it allows us to extend the results of this kind of study on
a finite system to much bigger systems. Contrary, the nontrivial
\textit{N-}dependent behavior of the thermodynamical variables and potentials
leads to a complication of the analysis.

In order to satisfy this scaling behavior for the global variables $E$, $N$,
$L$ and the Boltzmann entropy $S_{B}$, the local functions $\epsilon\left(
\mathbf{r}\right)  $, $\rho\left(  \mathbf{r}\right)  $, $\phi\left(
\mathbf{r}\right)  $ and $s\left(  \epsilon,\rho;\phi\right)  $ should be
scaled as follows:%

\begin{equation}
\left.  \rho\rightarrow\rho\left(  \alpha\right)  =\alpha^{\Lambda_{\rho}}%
\rho\right\}  \Rightarrow\left\{
\begin{array}
[c]{c}%
\phi\rightarrow\phi\left(  \alpha\right)  =\alpha^{\Lambda_{\phi}}\phi\\
\epsilon\rightarrow\epsilon\left(  \alpha\right)  =\alpha^{\Lambda_{e}%
}\epsilon\\
s\rightarrow s\left(  \alpha\right)  =\alpha^{\Lambda_{S}}s
\end{array}
\right\}  .
\end{equation}

Since the characteristic particles density behaves as $\rho_{c}\sim N/L^{3}$,
the scaling exponent for the particles density is $\Lambda_{\rho}%
=1-3\Lambda_{L}$. From the expression of the Newtonian potential (\ref{green})
is derived that its characteristic unit is $\phi_{c}\sim\rho_{c}L^{2}$, and
therefore, $\Lambda_{\phi}=1-\Lambda_{L}$. The energy scaling exponent is
equal to the scaling exponent of the total gravitational potential energy, so
that, $\Lambda_{E}=2-\Lambda_{L}$. The other scaling exponents are obtained by
using identical reasonings. All these scaling exponents depend on the scaling
exponent $\Lambda_{L}$ as follows:%

\begin{equation}%
\begin{array}
[c]{c}%
\Lambda_{\rho}=1-3\Lambda_{L}=\Lambda_{S},~\Lambda_{\phi}=1-\Lambda_{L},\\
\Lambda_{e}=2-4\Lambda_{L},~\Lambda_{E}=2-\Lambda_{L}.
\end{array}
~
\end{equation}

In order to satisfy these scaling laws is also \textit{necessary} that the
entropy density exhibits to the following scaling behavior:%

\begin{equation}
s\left(  \alpha^{\Lambda_{e}}\epsilon,\alpha^{\Lambda_{\rho}}\rho\right)
=\alpha^{\Lambda_{\rho}}s\left(  \epsilon,\rho\right)  ,
\end{equation}
This scaling property is satisfy if $s\left(  \epsilon,\rho\right)  $ obeys to
the following functional form:%

\begin{equation}
s\left(  \epsilon,\rho\right)  =\rho F\left(  \epsilon/\rho^{\eta}\right)  ,
\label{ent ff}%
\end{equation}
where $\eta=\Lambda_{e}/\Lambda_{\rho}$, being $F\left(  x\right)  $ an
arbitrary function. It is easy to show that the functional form (\ref{ent ff})
leads to the following relation between the pressure $p$ and the internal
energy density $\epsilon$:%

\begin{equation}
p=\gamma\epsilon, \label{gam}%
\end{equation}
where $\gamma=\eta-1$. There are some well-known Hamiltonian systems which
satisfy this kind of relation, as example, the system of nonrelativistic or
ultrarelativistic noninteracting particles, without matter if they obey to the
Boltzmann, Fermi-Dirac or Bose-Einstein Statistics. The scaling parameter
$\Lambda_{L}$ and $\Lambda_{E}$ are obtained from the parameter $\gamma$ as follows:%

\begin{equation}
\Lambda_{L}=\frac{\gamma-1}{3\gamma-1},~\,\Lambda_{E}=\frac{5\gamma-1}%
{3\gamma-1}.
\end{equation}

This result evidences that the power laws form for the self-similarity
conditions (\ref{self}) can be only satisfied by a reduced group of models
whose microscopic picture obeys to the relation\ (\ref{gam}). Since
$\gamma=\frac{2}{3}$ for the ideal gas of particles, the scaling exponents for
the Antonov problem \cite{antonov} and the selfgravitating fermions model
\cite{thir2,bilic} are given by $\Lambda_{E}=\frac{7}{3}$ and $\Lambda
_{L}=-\frac{1}{3}$, and therefore, they obey to the following thermodynamic limit:%

\begin{equation}
N\rightarrow\infty\text{, keeping constant }\frac{E}{N^{\frac{7}{3}}}\text{
and }LN^{\frac{1}{3}}\text{.} \label{thermodynamic limit}%
\end{equation}
This thermodynamic limit was established in the ref.\cite{chava} for the
self-gravitating nonrelativistic fermions by using other reasonings. Since the
selfgravitating relativistic gas and classic hard sphere model
\cite{oroson,stahl} do not obey to the relation (\ref{gam}), \textit{they do
not posses a scaling behavior with a\ power law form} (\ref{self}).

\subsection{Legendre formalism}

\label{formalism}For the sake of simplicity, let us consider firstly those
models exhibiting a power law thermodynamic limit (\ref{THL}). In such cases,
the steepest decent method leads to a thermodynamic formalism very analogue to
the one used for the extensive systems. The main contribution of the
functional integral (\ref{wm}) will come from the maxima of the functional
$\mathcal{L}\left[  \epsilon,\rho,\phi;z,z_{1},J\right]  $. This last one can
be rephrased as the Legendre functional, in which the entropy functional
$S\left[  \epsilon,\rho\right]  $ is maximized under the constrains of the
energy and particles number, where it is also taken into account the relation
between the Newtonian potential and the particles density. As elsewhere
discussed \cite{chava}, the existence of a multimodal functional
$\mathcal{L}\left[  \epsilon,\rho,\phi;z,z_{1},J\right]  $ for a given value
of the total energy $E$ tells us about the existence of several
quasi-equilibrium profiles, which could be related with the existence of phase
transitions \cite{chava,gross2}. When there is only one sharp peak, the
Boltzmann entropy can be appropriately estimated as follows:%

\begin{equation}
S_{B}\left(  E,N,L\right)  \simeq\underset{\epsilon,~\rho,~\phi}{\max}\left\{
\underset{\beta,~\mu,~j}{\min}\mathcal{L}\left[  \epsilon,\rho,\phi;\beta
,\mu,j\right]  \right\}  , \label{SS}%
\end{equation}
where $\beta$ and $\mu$ are the canonical parameters (real numbers), while $j$
is a real field which acts as a Lagrange multiplier for the relation between
the Newtonian potential and the particles density. The stationary conditions
lead to the following relations:%

\begin{equation}
\beta=\partial_{\epsilon}s\left(  \epsilon,\rho\right)  ,~\mu=-\frac{1}%
{2}\beta m\phi-\mathcal{G}\left[  j\right]  +\partial_{\rho}s\left(
\epsilon,\rho\right)  ,~j=\frac{1}{2}\beta m\rho, \label{bs}%
\end{equation}

\begin{equation}
N\left[  \rho\right]  =N,~~H\left[  \epsilon,\rho,\phi\right]  =E,~~\phi
=\mathcal{G}\left[  \rho\right]  . \label{norm}%
\end{equation}
The first relation of (\ref{bs}) states that the canonical parameter $\beta$
is \textit{constant} through all points of the system. Therefore, it is
convenient to use the Planck density $p_{c}\left(  \beta,\rho\right)
=\underset{\epsilon}{\max}\left[  \beta\epsilon-s\left(  \epsilon,\rho\right)
\right]  $ instead of the entropy density, which allows us to rewrite the
relations (\ref{bs}) as follows:%

\begin{equation}
\epsilon=\partial_{\beta}p_{c}\left(  \beta,\rho\right)  ,~\mu=-\beta
m\phi-\partial_{\rho}p_{c}\left(  \beta,\rho\right)  , \label{state functions}%
\end{equation}
where the field $j$ was eliminated and the identity%

\begin{equation}
\partial_{\rho}s\left(  \epsilon,\rho\right)  =-\partial_{\rho}p_{c}\left(
\beta,\rho\right)  ,
\end{equation}
was taken into account. The second relation in the equation
(\ref{state functions}) allows the obtaining of the state equation for the
particles density, $\rho=\rho\left(  \phi;\beta,\mu\right)  $. The
consideration of the constrains (\ref{norm}) lead to a self-consistent
integral equations system whose solution represents the most probable
equilibrium configuration of the astrophysical system for a given total energy
$E$.

Fluctuations around this equilibrium profile for the particles density $\rho$
could be estimated by using a Gaussian approximation as follows:%

\begin{equation}
\left(  \frac{\delta\rho}{\rho}\right)  ^{2}\simeq\left\vert \rho^{2}%
\frac{\delta^{2}}{\delta\rho^{2}}\mathcal{L}\right\vert _{eq}^{-1}.
\label{fluctua}%
\end{equation}
Since $\mathcal{L}$ grows proportional to $N$ when $N\gg1$, the relative
fluctuations depends on the system size as $\delta\rho/\rho\sim1/\sqrt{N}$, in
analogue form that the traditional systems do obey. Although all reasonings
have been made for astrophysical models with a power law thermodynamical limit
(\ref{THL}), the estimation of the relative fluctuations of the particles
density $\rho$ (\ref{fluctua}) evidences the general applicability of Legendre
formalism exposed above. This conclusion is straightforward followed from the
vanishing of the relative fluctuations of the particles density $\rho$ due to
the general linear growing of the functional $\mathcal{L}$ with the $N$ increasing.

This theoretical fact explains why several tools of the Traditional
Thermodynamics can be extended for the astrophysical objects in spite of their
nonextensive character. This conclusion has been elsewhere experimentally
confirmed by observations. For example, the evidence about the isothermal
character of the core in globular clusters and elliptical galaxies, where
these last ones display a quasi-universal luminosity profile described by de
Vaucouleur's $R^{1/4}$ law \cite{chava}; the presence of a dark matter halo
density profile decreasing as $r^{-2}$ at large distances in the spiral
galaxies, which is is also related with the isothermal distributions
\cite{bin}. However, in spite of the common applicability of several tools of
the standard thermodynamic formalism, the thermodynamical properties of
astrophysical systems turn to be very different from the ones exhibited by the
ordinary extensive systems due to the long-range character of the
gravitational interaction, which leads to the existence of some striking
phenomena, i.e. the gravitational collapse and the existence of a negative
heat capacity for certain energetic region
\cite{pad,kies,ruelle,lind,antonov,Lynden,Lynden1,thirring,gross,chava}.

\section{Geometrical considerations}

Is it only \textit{microcanonically} that could be performed a well-defined
description for the astrophysical systems? or, will there be also a
well-defined \textit{canonical-like} description? In our opinion, the answer
of the last question is \textit{yes}, which will be proved in the following subsections.

\subsection{Physics in the microcanonical ensemble is reparametrization
invariant.}

Let us consider certain functional $\varphi=\varphi\left(  E,N\right)  $ which
is a bijective function of the total energy $E$ (there is a bijective map
between $E$ and $\varphi$). In this case, the function $\varphi\left(
E,N\right)  $ is also an integral of motion for the system. Let us introduce
the microcanonical state density $\Omega_{\varphi}$:%

\begin{equation}
\Omega_{\varphi}\left(  \varphi\left(  E,N\right)  ,N\right)  =\int
\delta\left[  \varphi\left(  E,N\right)  -\varphi\left[  H_{N}\left(
X\right)  ,N\right]  \right]  dX.
\end{equation}
By using the identities:%

\begin{equation}
\delta\left[  \varphi\left(  E,N\right)  -\varphi\left(  H_{N},N\right)
\right]  =\left\vert \frac{\partial\varphi\left(  E,N\right)  }{\partial
E}\right\vert ^{-1}\delta\left(  E-H_{N}\right)  ,
\end{equation}
and%

\begin{equation}
\Omega_{\varphi}\left(  \varphi,N\right)  =\left\vert \frac{\partial
\varphi\left(  E,N\right)  }{\partial E}\right\vert ^{-1}\Omega_{E}\left(
E,N\right)  ,
\end{equation}
is straightforward derived the \textit{reparametrization invariance of the
microcanonical probabilistic distribution function:}%

\[
\frac{1}{\Omega_{\varphi}\left(  \varphi,N\right)  }\delta\left[
\varphi-\varphi\left(  H_{N},N\right)  \right]  =\frac{1}{\Omega_{E}\left(
E,N\right)  }\delta\left(  E-H_{N}\right)
\]

\begin{equation}
\omega_{M}\left(  X;\varphi,N\right)  =\omega_{M}\left(  X;E,N\right)  .
\end{equation}
From this property immediately follows the \textit{reparametrization
invariance of the microcanonical expectation values}:%

\[
\int F\left(  X\right)  \omega_{M}\left(  X;\varphi,N\right)  dX=\int F\left(
X\right)  \omega_{M}\left(  X;E,N\right)  dX,
\]

\begin{equation}
F_{M}\left(  \varphi,N\right)  =F_{M}\left(  E,N\right)  .
\end{equation}

The previous results point out the following conclusion: the microcanonical
description is \textit{equivalently} performed by using the thermodynamical
variables $\left(  E,N\right)  $ or $\left(  \varphi,N\right)  $. We say that
$\left(  E,N\right)  $ and $\left(  \varphi,N\right)  $ are two
\textit{equivalent representations} for the \textit{abstract space\ of the
microcanonical macroscopic description }$\Im_{N}$. The set of all those
representation changes among equivalent representations of $\Im_{N}$
constitute the\textit{\ group of diffeomorfisms or reparametrizations of }%
$\Im_{N}$, which is denoted by $Diff\left(  \Im_{N}\right)  .$ Thus, Physics
in microcanonical ensemble is\textit{\ invariant under the reparametrizations
changes of }$\Im_{N}$.

\subsection{Extensivity of the Planck thermodynamic potential.}

The analysis of the necessary conditions for the ensemble equivalence starts
from the consideration of the \textit{Laplace transformation} between the
microcanonical and the canonical-like partition functions, $\Omega_{\varphi
}\left(  \varphi,N\right)  $ and $Z_{\varphi}\left(  \beta_{\varphi},N\right)
$:%

\begin{equation}
Z_{\varphi}\left(  \beta_{\varphi},N\right)  =\int\exp\left(  -\beta
\varphi\right)  \Omega_{\varphi}\left(  \varphi,N\right)  d\varphi.
\label{Laplace}%
\end{equation}
Let us to show that it is always possible to choose a representation $\left(
\varphi,N\right)  $ of $\Im_{N}$ where the \textit{Planck potential}:%

\begin{equation}
\mathcal{P}_{\varphi}\left(  \beta_{\varphi},N\right)  =-\ln Z_{\varphi
}\left(  \beta_{\varphi},N\right)  ,
\end{equation}
is \textit{extensive }for a system with a power law self-similarity scaling
behavior (\ref{self}).

Let $W$ be the microcanonical phase-space accessible volume:%

\begin{equation}
W\left(  E,N\right)  =\Omega\left(  E,N\right)  \delta\epsilon,
\end{equation}
For continuous variables, $W$ is only well-defined after a coarsed grained
partition of phase space, which is the reason why it is considered a small
energy $\delta\epsilon$ in order to make $W$ dimensionless. However, during
the thermodynamic limit $N\rightarrow\infty$, we can make an arbitrary
selection of $\delta\epsilon$, whenever this last be small. In this case, the
microcanonical phase-space accessible volume $W$ appears as a
\textit{reparametrization invariant function}:%

\begin{equation}
W=\Omega_{E}\delta\epsilon=\Omega_{\varphi}\delta\varphi,
\end{equation}
where $\delta\epsilon$ and $\delta\varphi$ are very small in order to satisfy
the condition:%

\[
\delta\epsilon\simeq\left\vert \frac{\partial\varphi}{\partial E}\right\vert
^{-1}\delta\varphi.
\]
In this case, the Boltzmann entropy:
\begin{equation}
S_{B}=\ln W,
\end{equation}
becomes a \textit{scalar function}:%

\begin{equation}
S_{B}\left(  \varphi,N\right)  \equiv S_{B}\left(  E,N\right)  .
\end{equation}

Taking into account all the exposed above, the equation (\ref{Laplace}) can be
rewritten as follows:%

\begin{equation}
\exp\left[  -\mathcal{P}_{\varphi}\left(  \beta_{\varphi},N\right)  \right]
=\int\exp\left[  -\beta_{\varphi}\varphi+S_{B}\left(  \varphi,N\right)
\right]  \frac{d\varphi}{\delta\varphi}. \label{exp leg}%
\end{equation}
$S_{B}\left(  \varphi,N\right)  $ is extensive when the thermodynamic limit
(\ref{THL}) is carried out. The extensivity of the Planck potential
$\mathcal{P}_{\varphi}\left(  \beta_{\varphi},N\right)  $ when $N$ is tended
to infinity is guaranteed by choosing a functional $\varphi\left(  E,N\right)
$ which must be \textit{extensive} when the thermodynamic limit is performed .
This is a necessary condition for the ensemble equivalence. Although in the
microcanonical description any representation $\left(  \varphi,N\right)  $ of
$\Im_{N}$ can be considered as thermodynamic variables, in the canonical-like
description are \textit{only admissible} those representations satisfying the
extensity condition. This condition is arisen as the generalization of the
additivity condition of the traditional Thermodynamics.

The simplest choice for $\varphi\left(  E,N\right)  $, in accordance with the
thermodynamic limit (\ref{THL}), is given by the scaled energy $\mathcal{E}$:%

\begin{equation}
\mathcal{E}=E/N^{\Lambda_{E}-1}. \label{scaled energy}%
\end{equation}
In general, the extensivity of $\varphi\left(  E,N\right)  $ is ensured by
considering the following dependence:%

\begin{equation}
\varphi\left(  E,N\right)  =N\phi\left(  \mathcal{E}/N\right)  ,
\label{parametrization}%
\end{equation}
where $\phi\left(  \epsilon\right)  $ is a bijective function of $\epsilon$.
All those representation changes $\varphi\in Diff\left(  \Im_{N}\right)  $
preserving the extensivity of $\mathcal{E}$ in the thermodynamic limit
constitute a subgroup, which can be referred as the \textit{homogeneous
subgroup} $\mathcal{M}_{C}\subset Diff\left(  \Im_{N}\right)  $. All the
admissible representations for the canonical-like description are parametrized
by considering all transformations of $\mathcal{M}_{C}$ on the scaled energy
$\mathcal{E}$, the pair $\left(  \mathcal{E},\mathcal{M}_{C}\right)  $.

\subsection{Ensemble equivalence}

When $N$ is tended to infinity by using a representation $\mathcal{R}%
_{\varphi}\in\left(  \mathcal{E},\mathcal{M}_{C}\right)  $ in the equation
(\ref{exp leg}), the main contribution of this integral will come from the
maxima of the exponential argument. The equivalence between the microcanonical
and the generalized canonical ensemble will take place when there is only one
sharp peak. In such cases the validity of the \textit{Legendre transformation}
is ensured:%

\begin{equation}
\mathcal{P}_{\varphi}\left(  \beta_{\varphi},N\right)  =\min\left[
\beta_{\varphi}\varphi-S_{B}\left(  \varphi,N\right)  \right]  .
\end{equation}
The minimization leads to the conditions:%

\begin{equation}
\beta_{\varphi}=\frac{\partial}{\partial\varphi}S_{B}\left(  \varphi,N\right)
\text{, }K_{\varphi}=\frac{\partial^{2}}{\partial\varphi^{2}}S_{B}\left(
\varphi,N\right)  <0. \label{betak}%
\end{equation}

In this case, $K_{\varphi}=\partial_{\varphi}\beta_{\varphi}$ appears as a
generalization of the \textit{curvature tensor} of the microcanonical
thermostatistical theory of Gross \cite{gross2}. The \textit{topology} of this
tensor defines the \textit{ordering information} of the system by using a
microcanonical description. As usual, all those points of $\Im_{N}$ where
$K_{\varphi}>0$ can not be accessed by using the canonical-like description
\textit{in the }$\varphi$\textit{-representation, }$\mathcal{R}_{\varphi}%
$\textit{.} However, it can be easily shown that the sign of $K_{\varphi}$ is
\textit{representation dependent}.

Let $\mathcal{P}_{\varphi}^{m}\left(  \varphi,N\right)  $ be the
\textit{microcanonical Planck potential}:%

\begin{equation}
\mathcal{P}_{\varphi}^{m}\left(  \varphi,N\right)  =\varphi\frac{\partial
}{\partial\varphi}S\left(  \varphi,N\right)  -S_{B}\left(  \varphi,N\right)  .
\label{mic planck}%
\end{equation}
$\mathcal{P}_{\varphi}^{m}\left(  \varphi,N\right)  $ becomes in the canonical
Planck potential when the ensemble equivalence takes place in the
thermodynamic limit. Let us now consider two representations $\mathcal{R}%
_{\varphi_{1}}$ and $\mathcal{R}_{\varphi_{2}}$ of $\left(  \mathcal{E}%
,\mathcal{M}_{C}\right)  $, where the reparametrization change
$T_{1\rightarrow2}=\varphi_{2}o\varphi_{1}^{-1}\in\mathcal{M}_{C}$ :%

\begin{equation}
\varphi_{2}=\left(  \varphi_{2}o\varphi_{1}^{-1}\right)  \left(  \varphi
_{1}\right)  =\varphi_{2}\left(  \varphi_{1},N\right)  .
\end{equation}
Since $T_{1\rightarrow2}$ preserves the extensivity of $\varphi_{1}$, hence,
$\varphi_{2}\left(  \varphi_{1},N\right)  $ is a homogeneous function:%

\begin{equation}
\varphi_{2}\left(  \alpha\varphi_{1},\alpha N\right)  =\alpha\varphi
_{2}\left(  \varphi_{1},N\right)  ,
\end{equation}
and therefore, it obeys to the identity:%

\begin{equation}
\varphi_{2}=\varphi_{1}\frac{\partial\varphi_{2}}{\partial\varphi_{1}}%
+N\frac{\partial\varphi_{2}}{\partial N}. \label{homogen id}%
\end{equation}
Taking into consideration the definitions (\ref{betak}) and (\ref{mic planck}%
), as well as the identity (\ref{homogen id}), it is straightforward derived
the identities:%

\begin{equation}
\beta_{\varphi_{2}}=\left(  \frac{\partial\varphi_{1}}{\partial\varphi_{2}%
}\right)  \beta_{\varphi_{1}}, \label{beta_trans}%
\end{equation}

\begin{equation}
K_{\varphi_{2}}=\left(  \frac{\partial\varphi_{1}}{\partial\varphi_{2}%
}\right)  ^{2}\left[  K_{\varphi_{1}}+\frac{\partial}{\partial\varphi_{1}}%
\ln\left(  \frac{\partial\varphi_{1}}{\partial\varphi_{2}}\right)
\beta_{\varphi_{1}}\right]  , \label{curvature}%
\end{equation}

\begin{equation}
\mathcal{P}_{\varphi_{2}}^{m}\left(  \varphi_{2},N\right)  =\mathcal{P}%
_{\varphi_{1}}^{m}\left(  \varphi_{1},N\right)  +N\frac{\partial\varphi_{2}%
}{\partial N}\left(  \frac{\partial\varphi_{1}}{\partial\varphi_{2}}\right)
\beta_{\varphi_{1}}.
\end{equation}

It is evident from (\ref{curvature}) that the sign of $K_{\varphi}$ is
\textit{non invariant} under the reparametrization changes $\mathcal{M}_{C}$.
Note that $\beta_{\varphi}$ obeys the transformation rule for covariant
vectors in a \textit{Riemannian geometry}. However, $K_{\varphi}$ does not
obey the correct transformation rule of second-rank covariant tensors. This is
the reason why the curvature tensor is\textit{\ not a real tensor}, and the
ordering information which it contains is non invariant to under the
reparametrization changes of $\mathcal{M}_{C}$. A question arises: is it
possible to choose a representation $\mathcal{R}_{\varphi}$ in which the
ensemble equivalence takes place? The answer is \textit{yes}, and we will
prove it now.

As already said, when it is used the scaled energy representation
$\mathcal{R}_{\mathcal{E}}=\left(  \mathcal{E},N\right)  $
(\ref{scaled energy}), there is an energetic region where the astrophysical
systems exhibits a negative heat capacity: $K_{\mathcal{E}}$ $>0$. Let us
consider another representation $\mathcal{R}_{\Phi}=\left(  \Phi,N\right)  $
which can be parametrized by a bijective function $\varphi\left(
\epsilon\right)  $ as follows:%

\begin{equation}
\Phi=N\varphi\left(  \epsilon\right)  \text{ with }\epsilon=\mathcal{E}/N.
\end{equation}
In order to disregard the \textit{N}-dependence, let us work with the
Boltzmann entropy per particle in the thermodynamic limit:%

\begin{equation}
s\left(  \epsilon\right)  =\underset{N\rightarrow\infty}{\lim}\frac
{S_{B}\left(  \mathcal{E},N\right)  }{N}=s\left(  \varphi\right)  ,
\end{equation}
as well the \textit{N}-independent thermodynamic variables $\epsilon$ and
$\varphi=\Phi/N$. In this case, the canonical parameter $\beta$ and the
\textit{N}-independent curvature $k$ are denoted by:%

\begin{equation}
\beta_{\epsilon}=\partial_{\epsilon}s,\text{ }k_{\epsilon}=\partial_{\epsilon
}\beta_{\epsilon},
\end{equation}
in the $\mathcal{R}_{\mathcal{E}}$ representation, or%

\begin{equation}
\beta_{\varphi}=\partial_{\varphi}s,\text{ }k_{\varphi}=\partial_{\varphi
}\beta_{\varphi},
\end{equation}
in the $\mathcal{R}_{\Phi}$ representation. The notation $\partial_{\epsilon
}s$ is equivalent to $ds/d\epsilon$. The expression (\ref{curvature}) can be
rewritten in this case as follows:%

\[
\partial_{\varphi}\beta_{\varphi}\left(  \varphi\right)  =\left(
\partial_{\epsilon}\varphi\left(  \epsilon\right)  \right)  ^{-2}\left[
\partial_{\epsilon}\beta_{\epsilon}\left(  \epsilon\right)  -\beta_{\epsilon
}\left(  \epsilon\right)  \partial_{\epsilon}\ln\left(  \partial_{\epsilon
}\varphi\left(  \epsilon\right)  \right)  \right]
\]%
\begin{equation}
\partial_{\varphi}\beta_{\varphi}\equiv-\left(  \partial_{\epsilon}%
\varphi\left(  \epsilon\right)  \right)  ^{-2}a\left(  \epsilon\right)  ,
\end{equation}
where the function $a\left(  \epsilon\right)  $:%

\begin{equation}
\partial_{\epsilon}\beta_{\epsilon}-\beta_{\epsilon}\partial_{\epsilon}%
\ln\left(  \partial_{\epsilon}\varphi\left(  \epsilon\right)  \right)
=-a\left(  \epsilon\right)  . \label{a def}%
\end{equation}
\textit{should be positive in order to ensure the ensemble equivalence in the}
$\mathcal{R}_{\Phi}$ \textit{representation. }Direct integration of
(\ref{a def}) yields:%

\begin{equation}
\partial_{\epsilon}\varphi\left(  \epsilon\right)  =C\beta_{\epsilon}\left(
\epsilon\right)  \exp\left(  \int\frac{a\left(  \epsilon\right)  }%
{\beta_{\epsilon}\left(  \epsilon\right)  }d\epsilon\right)  ,
\end{equation}
where $C$ is a positive constant which could be set as unity. It is easy to
see that a convenient choice for the function $a\left(  \epsilon\right)  $,
from the theoretical viewpoint, could be given by:%

\begin{equation}
a\left(  \epsilon\right)  =\left\{
\begin{tabular}
[c]{lll}%
$-k_{\epsilon}\left(  \epsilon\right)  $ & if & $k_{\epsilon}\left(
\epsilon\right)  <0,$\\
$\beta_{\epsilon}^{2}\left(  \epsilon\right)  =\beta_{c}^{2}\sim const$ \ \  &
if & $k_{\epsilon}\left(  \epsilon\right)  =0,$\\
$k_{\epsilon}\left(  \epsilon\right)  $ & if & $k_{\epsilon}\left(
\epsilon\right)  >0,$%
\end{tabular}
\ \ \right\vert
\end{equation}
which ensures the positivity of $a\left(  \epsilon\right)  $. Thus,
$\partial_{\epsilon}\varphi\left(  \epsilon\right)  $ is given by:%

\begin{equation}
\partial_{\epsilon}\varphi\left(  \epsilon\right)  =\left\{
\begin{tabular}
[c]{lll}%
$1$ & if & $k_{\epsilon}\left(  \epsilon\right)  <0,$\\
$\beta_{c}\exp\left(  \beta_{c}\epsilon\right)  $ & if & $k_{\epsilon}\left(
\epsilon\right)  =0,$\\
$\beta_{o}^{2}\left(  \epsilon\right)  $ & if & $k_{\epsilon}\left(
\epsilon\right)  >0.$%
\end{tabular}
\ \ \right\vert \label{dgk}%
\end{equation}

According to (\ref{dgk}), $\varphi\left(  \epsilon\right)  $ is an piecewise
monotonic function of $\epsilon$, and therefore, there is a bijective
correspondence between $\epsilon$ and $\varphi\left(  \epsilon\right)  $.
Hence, $\epsilon$ and $\varphi$ are equivalent representations, where
$k_{\varphi}=\partial_{\varphi}\beta_{\varphi}$ is negative; and therefore,
the canonical-like description which uses the representation $\mathcal{R}%
_{\Phi}=\left(  \Phi,N\right)  $ is \textit{equivalent }to the microcanonical
description in the thermodynamic limit\textit{. It was shown in this way that
it is always possible to choose a representation where the ensemble
equivalence is guaranteed. }

\subsection{An example: the Antonov problem.}

As example, let us consider the classical Antonov problem \cite{antonov}: the
self-gravitating gas interacting by means of the Newtonian potential, which is
enclosed with a spherical rigid container of radio $R$. Considering that
$G=M=R=\hbar=1$, the microcanonical description of this model system is
derived from the Planck density:%

\begin{equation}
p_{c}\left(  \beta,\rho\right)  =\frac{3}{2}\rho\ln\left(  2\pi\beta\right)
+\rho\ln\rho-\rho,
\end{equation}
by using the thermodynamic formalism presented in subsection \ref{formalism}.
The internal energy density $\epsilon$ and particles density $\rho$ are given by:%

\begin{equation}
\epsilon=\frac{3}{2\beta}\rho,~\rho=C\exp\left(  \Phi\right)  ,
\end{equation}
where $C=\left(  2\pi\beta\right)  ^{-\frac{3}{2}}\exp\left(  -\mu\right)  $
and $\Phi=-\beta\phi,$~being $\phi$ the Newtonian potential. The numerical
study is carried out by solving the Poisson-Boltzmann equation:%

\begin{equation}
\Delta\Psi=-4\pi\exp\left(  \Psi\right)  ,
\end{equation}
where $\Psi=\Phi+\ln\left(  \beta C\right)  \equiv\ln\left(  \beta\rho\right)
$, imposing the following conditions at the origin:%

\begin{equation}
\Psi\left(  0\right)  =\psi,\Psi^{\prime}\left(  0\right)  =0,
\end{equation}
being $\psi=\ln\left(  \beta\rho_{0}\right)  $ an integration parameter
related with the central density $\rho_{0}$ of the system. The canonical
parameters $\beta$ and $\mu$ are obtained through the relations:%

\begin{equation}
\Psi\left(  1\right)  =\beta+\ln\left(  \beta C\right)  ,\Psi^{\prime}\left(
1\right)  =-\beta,
\end{equation}
while the entropy and the total energy are obtained from the relations
(\ref{SS}) and (\ref{norm}) respectively. It is easy to see that \textit{all
the thermodynamical variables and potentials are obtained as functions of the
parameter} $\psi$.%

\begin{figure}
[h]
\begin{center}
\includegraphics[
height=2.6238in,
width=2.7726in
]%
{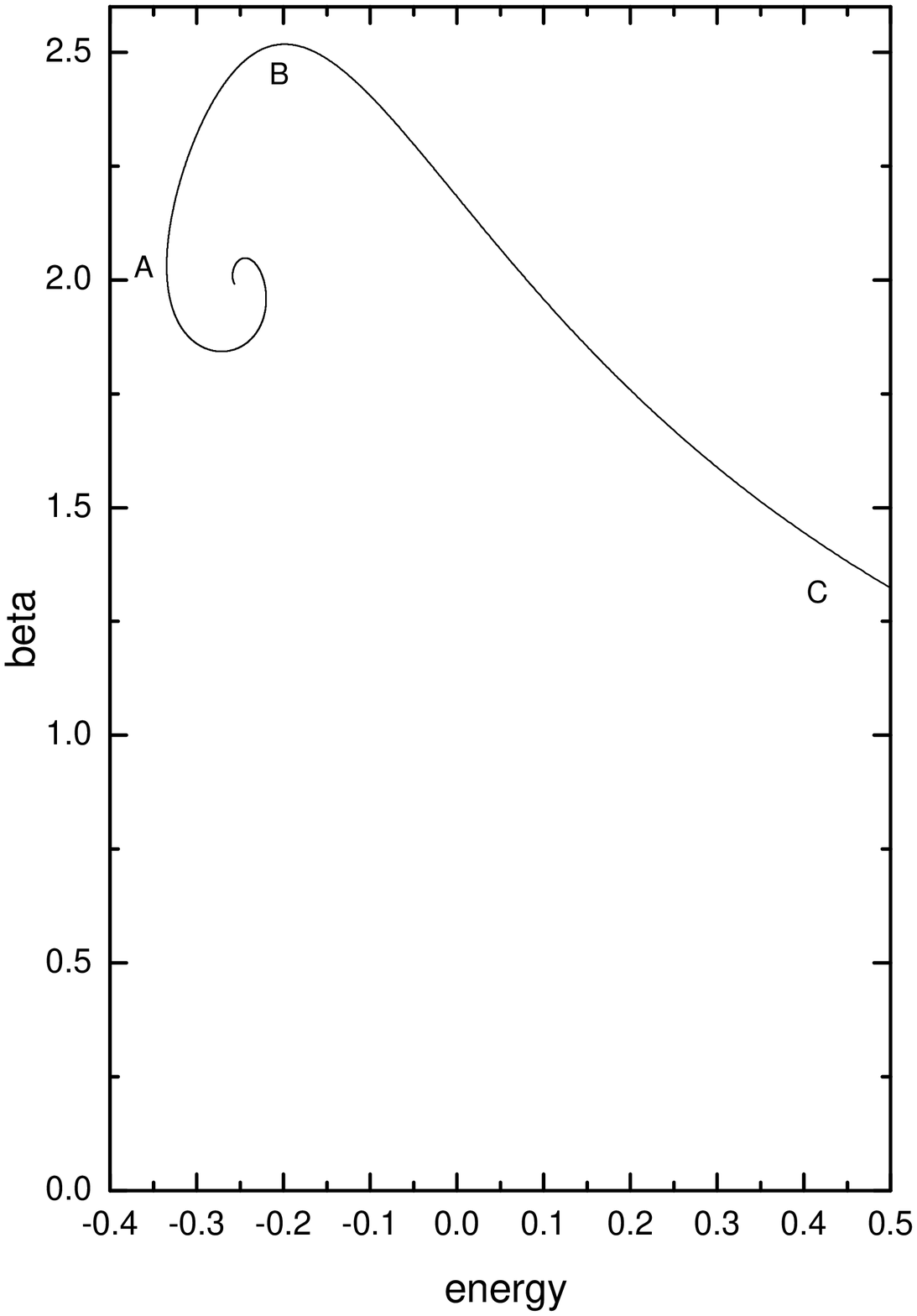}%
\caption{Caloric curve for the Antonov problem. The states beloging to the
interval $AB$ exhibit a negative specific heat, and therefore, they are not
accessible by using the canonical ensemble.}%
\label{espiral2}%
\end{center}
\end{figure}

Figure (\ref{espiral2}) shows the well-known caloric curve of the Antonov
problem. The points of superior branch $ABC$ correspond to the equilibrium
states, while the others represent unstable saddle points. No equilibrium
states are found for energies $\epsilon<-0.335$, which is related with the
existence of the gravitational collapse. The equilibrium states belonging to
the energetic interval $-0.335<\epsilon<-0.198$ exhibit a negative specific
heat, and therefore, they are not accessible by using the canonical ensemble.

According to all the exposed in subsection above, the Antonov problem obeys to
the thermodynamic limit (\ref{thermodynamic limit}), and therefore, we are
able to choose an appropriate representation for the microcanonical variables
in order to guarantee that the canonical-like ensemble associated with this
new representation be equivalent to the microcanonical one.%

\begin{figure}
[h]
\begin{center}
\includegraphics[
height=2.7484in,
width=2.7614in
]%
{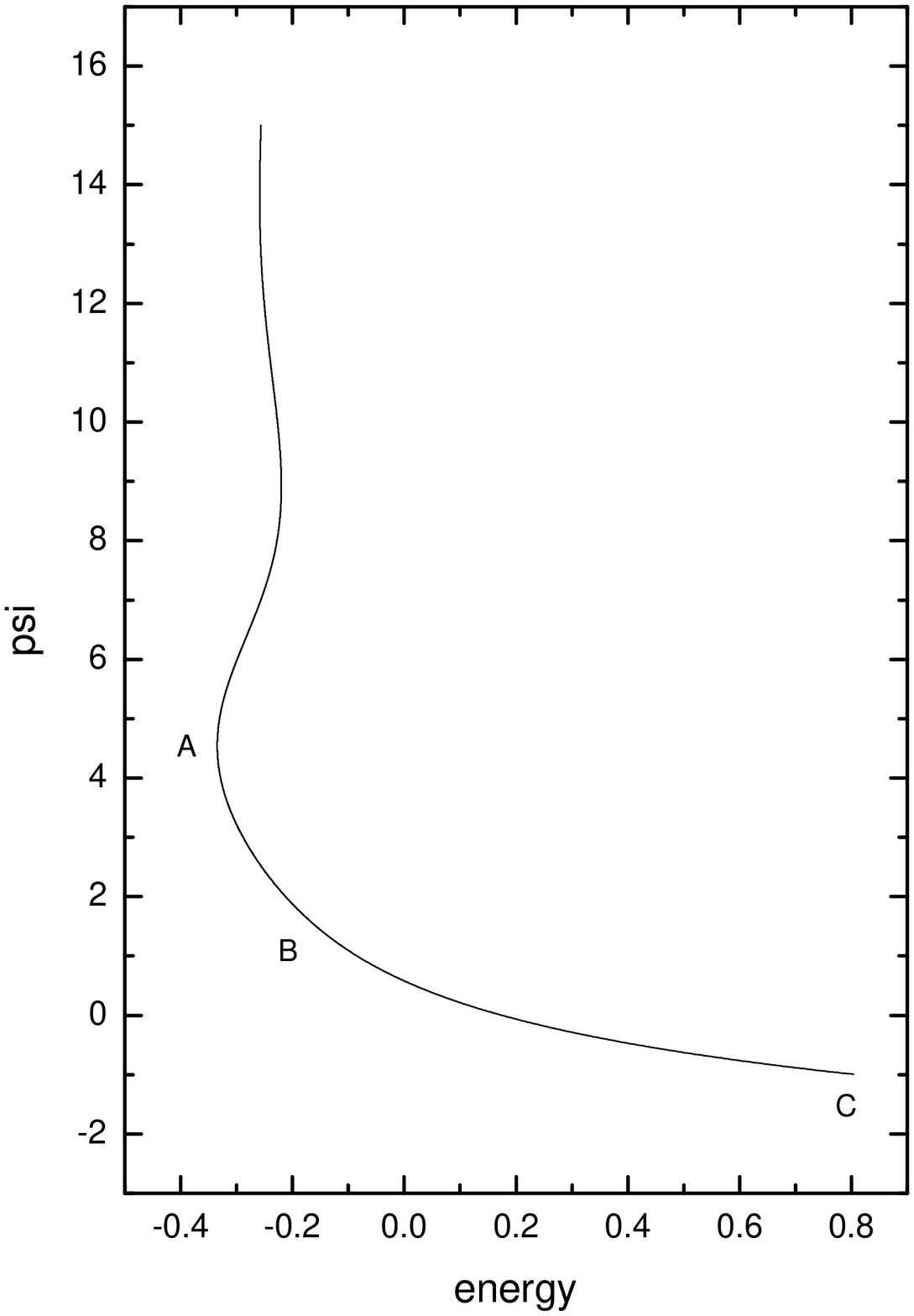}%
\caption{The curve $\psi-\epsilon$. There is a univocal correspondence between
these variables for all equilibrium states ($ABC$ branch).}%
\label{density}%
\end{center}
\end{figure}

\begin{figure}
[h]
\begin{center}
\includegraphics[
height=2.6437in,
width=2.6481in
]%
{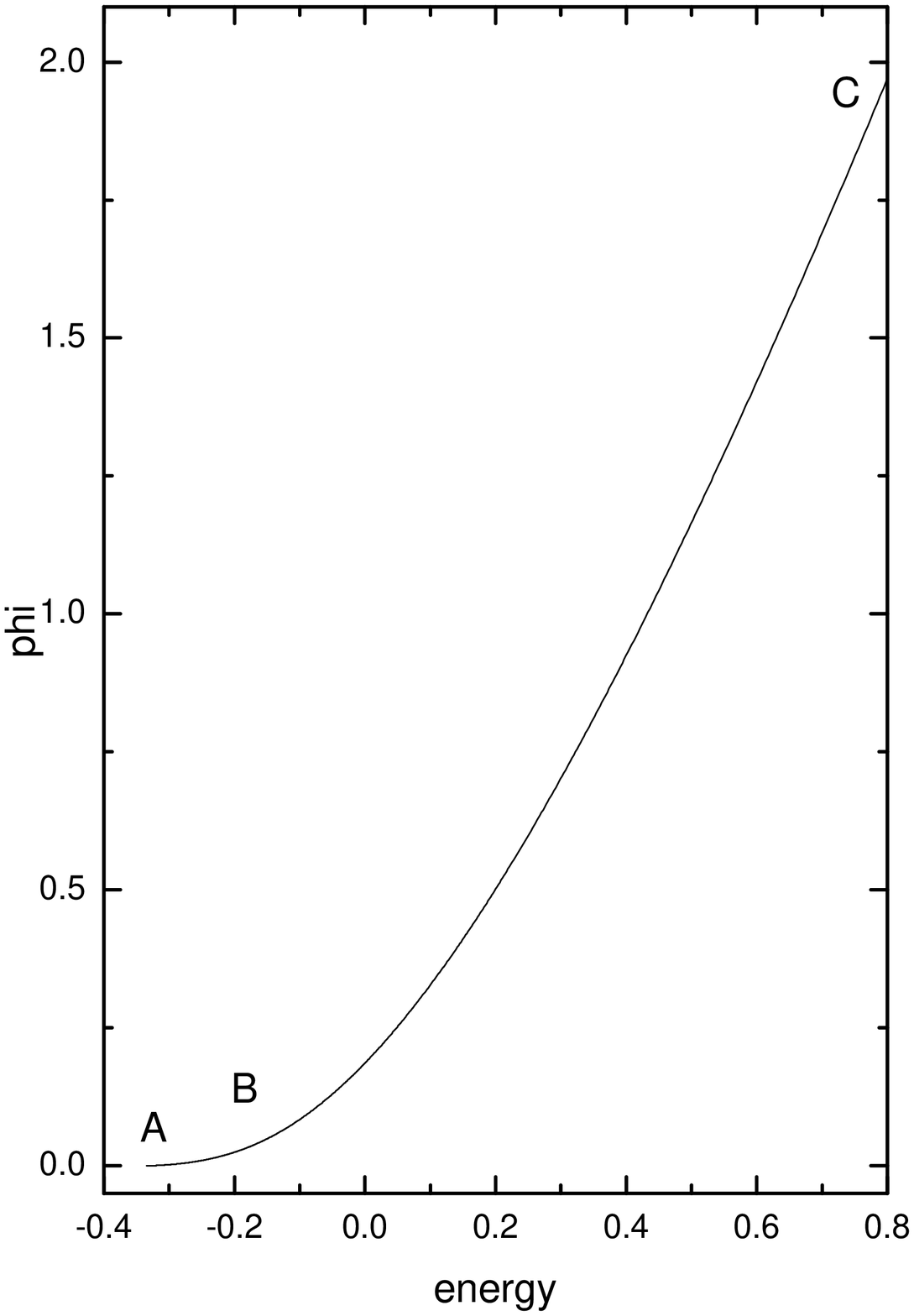}%
\caption{The map: $\varphi:\epsilon\rightarrow\varphi\left(  \epsilon\right)
$. The bijective character of this transformation ensures the invariance of
the microcanonical description.}%
\label{map}%
\end{center}
\end{figure}

\begin{figure}
[h]
\begin{center}
\includegraphics[
height=2.687in,
width=2.7596in
]%
{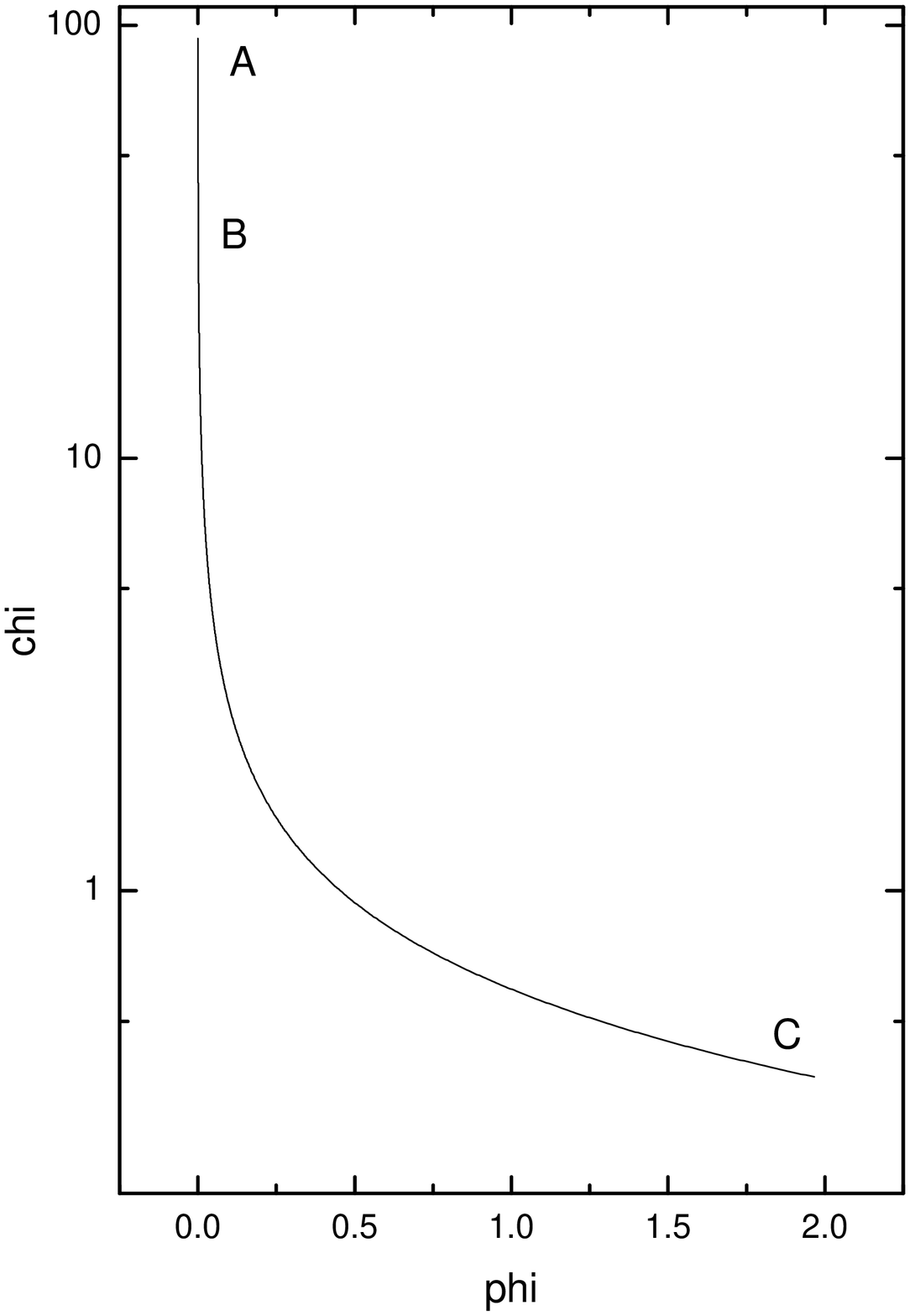}%
\caption{The \textquotedblleft caloric curve\textquotedblright\ $\chi-\varphi
$. This figure reveals the equivalence between the canonical-like ensemble
with the microcanonical ones in the $\varphi$-representation.}%
\label{new_cal}%
\end{center}
\end{figure}

\begin{figure}
[h]
\begin{center}
\includegraphics[
height=2.7069in,
width=2.7371in
]%
{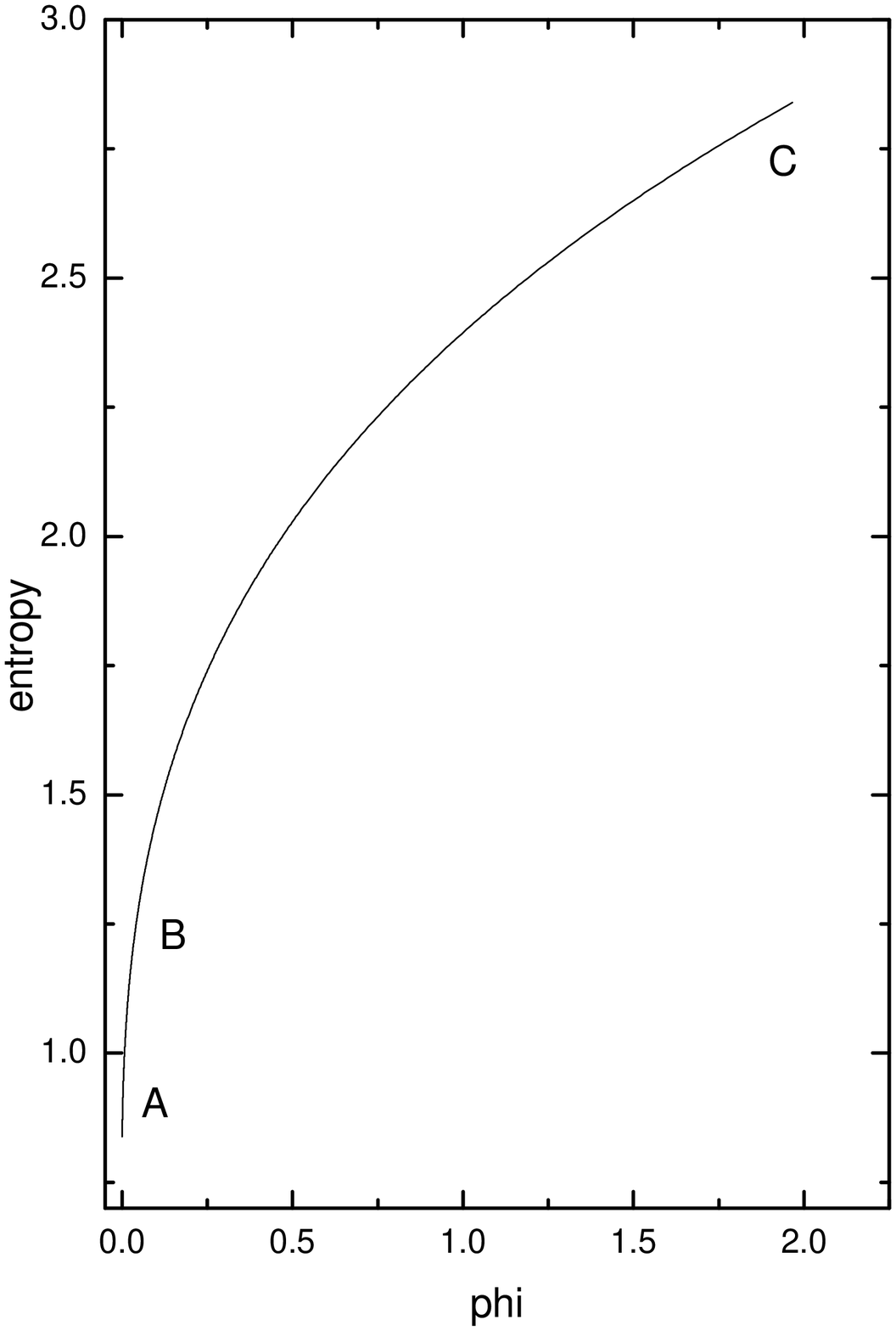}%
\caption{The entropy function in the $\varphi$-representation. It can be note
the concavity of this thermodynamical potential.}%
\label{new_ent}%
\end{center}
\end{figure}

Figure (\ref{density}) shows the dependence between the thermodynamical
variables $\epsilon$ and $\psi$. Disregarding the unstable branch, it is easy
to see that there is a univocal correspondence between these variables for all
equilibrium states. We can select an appropriate representation $\varphi
=\varphi\left(  \epsilon\right)  $ for the microcanonical description by
rephrasing $\psi$, or more exactly, $\chi=\exp\left(  \psi\right)  $ as a
\textit{canonical-like parameter}. The map $\epsilon\rightarrow\varphi\left(
\epsilon\right)  $ can be established by using the transformation rule
(\ref{beta_trans}) for the canonical parameters:%

\begin{equation}
\chi=\exp\left(  \psi\right)  =\frac{\partial\epsilon}{\partial\varphi}%
\beta\Rightarrow\frac{d\varphi\left(  \psi\right)  }{d\psi}=\beta\left(
\psi\right)  \exp\left(  -\psi\right)  \frac{d\epsilon\left(  \psi\right)
}{d\psi}.
\end{equation}
This map is shown at figure (\ref{map}). The bijective character of this
representation change assures the reparametrization invariance of Physics in
the microcanonical ensemble. Figures (\ref{new_cal}) and (\ref{new_ent}) show
the \textquotedblleft caloric curve\textquotedblright\ $\chi-\varphi$ and the
entropy function $s-\varphi$ in the $\varphi$ representation for the
microcanonical variable $\epsilon$. The first shows the equivalence between
the canonical-like ensemble with the microcanonical ones, while the second one
confirms the concavity of the entropy function in this representation.%

%

%

Thus, the physical observables obtained from the canonical-like probabilistic
distribution function:%

\begin{equation}
\omega_{c}\left(  E~;\chi,N\right)  =\frac{1}{\mathcal{Z}_{\varphi}\left(
\chi,N\right)  }\exp\left[  -N\chi\cdot\varphi\left(  E~/N^{\frac{7}{3}%
}\right)  \right]  , \label{chi-pdf}%
\end{equation}
are equal to those obtained by using the microcanonical description in the
thermodynamic limit (\ref{thermodynamic limit}). The numerical solution of
this model exposed above was obtained by considering fixed the parameter
$\psi=\ln\chi$, and this supposition allows us to determine all the
thermodynamical observables of this model system. Therefore, we are able to
affirm that this numerical solution was obtained by using the $\chi
$-\textit{canonical-like} description.

It is remarkable that this generalized ensemble allows us to access to all
equilibrium states of the model, even in the cases in which these last ones
exhibit a \textit{negative }specific heat. This is an important difference
regarding to the usual canonical ensemble. Although the similar appearance
between these two descriptions, the simple representation change of the
microcanonical variables (compatible with the scaling laws of the
thermodynamical variables and potentials in the thermodynamic limit) allows
the extension of a canonical-like description to those cases in which the
usual Gibbs' canonical one fails.

\section{Concluding remarks}

We have reconsidered some aspects about the microcanonical mean field
description for astrophysical Hamiltonian systems. Particularly, we derived a
formalism which takes into account those cases is which the particles also
interact by means of short-range forces. This general framework allows us to
perform the analysis of the necessary conditions for the existence of a power
law self-similarity behavior of the microcanonical variables of the form
(\ref{self}), which is an extension of extensivity of the traditional systems.
This study revealed that this property could be only satisfied by a reduced
set of models where the nature of the short-range interactions leads to the
relation $p=\gamma\epsilon$ between the pressure $p$ and the internal energy
per volume $\epsilon$, being $\gamma$ certain constant.

As already shown, the consideration of the reparametrization invariance of the
microcanonical ensemble allows us to extend the applicability of a
canonical-like description to those situations where the standard
Thermodynamics based on the canonical ensemble is not able to describe, for
instance, when the systems exhibit a negative specific heat. This possibility
was illustrated in this paper by reconsidering the classical isothermal model
of Antonov \cite{antonov}.

The physical significance of such generalized canonical-like ensemble relies
on the reparametrization invariance of Physics in the microcanonical ensemble
as well as the equivalence of these descriptions by the consideration of an
appropriate thermodynamic limit. It is important to remak that such
thermodynamic limit is derived from the scaling properties of the
thermodynamical variables and potentials of a system with a large number of
degrees of freedom, and therefore, the same one can not be arbitrarily introduced.

An interesting application field of this geometrical ideas could be in the
Montecarlo methods based on the canonical weight $\exp\left(  -\beta E\right)
$. It is well-known that these numerical methods are more simple than those
based on the microcanonical ensemble, but the first ones diverge when the
ensemble equivalence is not ensured. Therefore, the consideration of a
generalized canonical-like weight of the form $\exp\left[  -N\chi\cdot
\varphi\left(  E/N^{\Lambda_{E}}\right)  \right]  $ (where $\Lambda_{E}$ is
the scaling exponent of energy in the thermodynamic limit $N\rightarrow\infty
$), could guarantee the convergence of such methods for all the energetic range.

\end{document}